\documentstyle[preprint,revtex,eqsecnum]{aps}
\begin{document}
\preprint{IFT-P.027/93}
\preprint{August 1993}
\preprint{hep-ph/9308282}

\begin{title}
Does $K_L-K_S$ mass difference constraints or \\ claims
new physics beyond the Standard Model?
\end{title}
\author{ F. Pisano and V. Pleitez}
\begin{instit}
Instituto de F\'\i sica Te\'orica\\
Universidade Estadual Paulista\\
Rua Pamplona, 145\\
01405-900-- S\~ao Paulo, SP\\
Brazil
\end{instit}
\begin{abstract}
The ratio $\Delta m_K/m_K$ within the standard model with 3
generations is calculated as a function of the CP
nonconserving phase $\delta_{13}$ and the quark masses $m_c,m_t$ assuming
the current values of the Cabibbo-Kobayashi-Maskawa mixing angles. We
have found that varying $\delta_{13}$ and $m_c$ within the allowed range,
not all the values for the top quark mass fit the experimental value
for that ratio.
\end{abstract}

The absence of $\Delta S\not=0$ neutral currents led to the so called
Glashow-Iliopoulos-Maiani (GIM) mechanism~\cite{gim} in the
context of the $SU(2)_L\otimes U(1)_Y$ gauge theory~\cite{wsg} with
four quarks (two generations). The GIM mechanism is also implemented
in the 3 generations case~\cite{km,gw}.
However, the natural (independent of mixing angles) flavor conservation
is a characteristic of the standard model without heavy
quarks~\cite{gw}, i.e.,  $m^2_q/M^2_W\ll1$.
On the other hand, if there are quarks with mass
$m^2_q/M^2_W\approx 1$ or greater, the requirements for the natural
flavor conservation in the neutral currents to order $\alpha G_F$
break down, and it is necessary to impose the restriction that the
mixing angles between ordinary and superheavy quark sectors must be
very small~\cite{poggio}.

It is well known that the $K_L$-$K_S$ mass difference was used to
estimate the mass of the $c$ quark even before its
discovery~\cite{gl}. Since then, this was used to argue that any
additional contribution to this mass difference coming from new
particles, usually present in models beyond the standard model,
cannot be much bigger than the contribution of the $c$ quark. This
means that the observation of even a tiny flavor changing neutral
currents (FCNC) effect would imply new physics beyond the standard
model. Since then, it has been mandatory to study the top quark
contribution to rare processes in kaon ~\cite{res} and beauty
{}~\cite{vb} mesons. In particular the $K_L$-$K_S$ mass difference
($\Delta m_K=m_{K_L}-m_{K_S}$) and $CP$
non--conservation were used to limit the range of the
Cabibbo-Kobayashi-Maskawa (CKM) mixing angles $\theta_{2,3}$
and phase parameter $\delta_{13}$. These processes in addition with
$B$-$\bar B$ mixing, and the ratio $\Gamma(W)/\Gamma(Z)$ extracted from
$p\bar p$ colliders were used to fit the top quark mass~\cite{fh}
\begin{equation}
35\,\mbox{GeV}/c^2\lower.7ex\hbox{$\;\stackrel{\textstyle<}{\sim}\;$}
m_t\lower.7ex\hbox{$\;\stackrel{\textstyle<}{\sim}\;$}55\,\mbox{GeV}/c^2.
\label{topfit}
\end{equation}

At present, however, we know that~\cite{cdf2}
\begin{equation}
m_t>108\,\mbox{GeV}/c^2.
\label{mt}
\end{equation}
On the other hand, from radiative corrections it was obtained
that~\cite{pl}
\begin{equation}
m_t \leq 200\,\mbox{GeV}/c^2,
\label{mt2}
\end{equation}
assuming that the Higgs bosons mass $m_H$ is not larger than
$ 1$ TeV. With such a value of $m_t$ we see, as it was stressed in
Ref.~\cite{poggio}, that the suppression of FCNC
to order $\alpha G$ requires that the mixing angles between
the $t$ quark and the lighter quarks must be very small.

In this work we will show that either: {\em i)}
we do not know the correct CKM
parameters yet, or {\em ii)} not all values of $m_t$ can accommodate the
$K_L$-$K_S$ mass difference. The last point is probably valid for
other rare processes but we will not address this issue here~\cite{ajb1}.

We will assume that CKM mixing
angles have the present values: $s_{12}=0.218$ to $0.224$,
$s_{23}=0.032$ to
$0.054$, and $s_{13}=0.002$ to $0.0077$~\cite{pdg}, next we calculate
$\Delta m_K/m_K$ for several values of $\delta_{13}$, $m_c$ and $m_t$.
We did this not only for
the sake of simplicity but because, as it is usually believed,
whereas the tree level decays are sensitive mainly to the hierarchy
of weak couplings, the radiative corrections in decays and
transitions depend mainly on the hierarchy of quark masses. In the
box diagrams usually the $m_c$ entering in the calculation is the
running $c$ quark mass. Here we will allow this mass to run upon the range
$1.3$--$1.7\,\mbox{GeV}/c^2$ which is consistent with the
experimental value~\cite{pdg}.

In fact, it has been shown that $m_t$ is almost independent of the
CKM parameters~\cite{ajb2,cudell}. It means that only when
the heavy quark masses become determined from scattering processes
and/or spectroscopy we could look for their effects in rare decays,
mixing and $CP$ violation in $K$-, $B$- and $D$- mesons.

Of course, this is not easy to be implemented because the top quark
has not been discovered yet but it is known that its mass must satisfy
the lower bound given in Eq.~(\ref{mt}).
However, in the future, we expect that the
top quark mass can be well determined in processes different from
those in which the weak couplings are determined.
For the reasons above we will assume, as we said before, that we
already know the mixing angles in the CMK matrix but we allow to
vary the phase $\delta_{13}=\delta$~\cite{fn} and $m_t$ for a given
mass of the $c$ quark. With these free parameters we study the
contributions to the $K_L$-$K_S$ mass difference. In particular, we
will compare this quantity with its experimental value.

Neglecting long distance contributions, one has that the ratio is
given by~\cite{japa,ajb3}
\begin{equation}
\frac{\Delta
m_K}{m_K}=\frac{2}{3}\kappa\,B_Kf^2_K\frac{G_F}{\sqrt2}\, \tilde
E(x_i,x_j),
\label{delta}
\end{equation}
where $\kappa=\alpha(4\pi\sin^2\theta_W)^{-1}$ and
\begin{eqnarray}
\tilde E(x_i,x_j)&=&\left[(Re\,\lambda_c)^2-(Im\,\lambda_c)^2\right]
E(x_c,x_c)\eta_1+\left[(Re\,\lambda_t)^2-(Im\,\lambda_t)^2
\right]E(x_t,x_t)\eta_2
\nonumber \\ &
&\mbox{}+\left[(Re\,\lambda^2_c)^2(Re\,\lambda^2_t)^2-(Im\,\lambda_c)^2
(Im\,\lambda_t)^2\right]E(x_c,x_t)\eta_3,
\label{e}
\end{eqnarray}
where $\lambda_i=V^*_{is}V_{id}$, $V_{ij}$ are the CKM matrix
elements in the parametrization of
Maiani~\cite{pdg,maiani} and $x_i=m_i^2/M^2_W$. The QCD corrections,
which depend weakly on the top mass, are in the $\eta_i$
coefficients~\cite{gilman}.

First we will use $\eta_i=1$,
$i=1,2,3$. Considering the free quark model is interesting because it
is the cleaner calculated part. It is important to recall, that it
was in the free-quark case that the value of $m_c$ was predicted. At
the end of this work we will comment the QCD corrections.

In Eq.~(\ref{delta}) the
``bag'' parameter $B_K$ depends on the calculation of the amplitude of the
transition $\bar K^0\leftrightarrow K^0$. We will use
$B_K=1$ and $f_K=161\pm1$ MeV for the kaon decay constant
{}~\cite{ajb2}.

The functions $E(x_i,x_j)$ in Eq.~(\ref{e}) were calculated by Inami
and Lim~\cite{japa} and by Buras~\cite{ajb3} and they are given by
\begin{equation}
E(x_i,x_i)=x_i[\frac{1}{4}+\frac{9}{4}(1-x_i)^{-1}-\frac{3}{2}
(1-x_i)^{-2}]+\frac{3}{2}[x_i/(x_i-1)^2]^{3}\ln x_i,
\label{xi}
\end{equation}

\begin{eqnarray}
E(x_i,x_j)&=&x_ix_j\{(x_i-x_j)^{-1}[\frac{1}{4}+\frac{3}{2}(1-x_j)^{-1}
-\frac{3}{4}(1-x_j)^{-2} ]\ln x_i\nonumber \\
& &+x_i\leftrightarrow x_j -\frac{3}{4}[(1-x_j)(1-x_i)]^{-1}.
\}
\label{xij}
\end{eqnarray}
The $u$ quark contribution in Eqs.~(\ref{delta}) is rearranged into
the heavy quark contributions~\cite{japa}.

We use the CMK matrix elements given in Ref.~\cite{pdg} which
correspond to the central values $\sin\theta_{12}=0.221$,
$\sin\theta_{23}=0.043$ and $\sin\theta_{13}=0.036$. We recall that
at present it is an open question what the form of the unitary
triangle is, i.e, if $\pi/2\leq\delta\leq\pi$ or
$0\leq\delta\leq\pi/2$. For this reason, we treat this phase as a
free parameter.

In Fig. 1 we give the calculated values
of $\Delta m_K/m_K$ in terms of the top quark mass for different
values of the charm quark mass and the $\delta$ parameter. The error
in $f_K$ is not shown in the figure.
The experimental value $\Delta
m_K/m_K=7.08\times10^{-15}$~\cite{pdg} is shown as a horizontal
dashed line.

We see from Fig. 1 that only the following values make the
standard model consistent with the experimental value of $\Delta
m_K/m_K$:
\begin{equation}
m_c=1.6\,\, \mbox{GeV}/c^2,\quad m_t\simeq 157\,\,\mbox{GeV}/c^2,\quad
\delta \lower.7ex\hbox{$\;\stackrel{\textstyle<}{\sim}\;$}\pi.
\label{values}
\end{equation}
For lower (higher) values of the quark masses there is a deficit
(excess) on the $\Delta m_K/m_K$. For instance, for
$m_c=1.3\,\mbox{GeV}/c^2$,
$\delta=\pi/4$ and $m_t\approx110\,\mbox{GeV}/c^2 $ we have $\Delta
m_K/m_K\simeq 0.62(\Delta
m_K/m_K)^{exp}$, a deficit of $\sim38\%$, whereas for
$m_c=1.7\,\,\mbox{GeV}/c^2$,
$\delta=\pi$ and $m_t\approx110\,\,\mbox{GeV}/c^2$ we have $\Delta
m_K/m_K\simeq 1.08(\Delta
m_K/m_K)^{exp}$ that is, an excess of about $8\%$.

We have repeated our analysis varying the Cabibbo-Kobayashi-Maskawa
parameters within the allowed range~\cite{pdg}, however, in this case the
numerical results almost do not vary and
the main feature remains: {\em within the standard model scenario for
all current values allowed for the CKM
parameters, only the masses of the quarks $m_c$ and $m_t$ and
$\delta$ given in Eq.~(\ref{values}) fit the experimental data of the
$K_L$-$K_S$ mass
difference.}

Of course, strong interaction corrections short ($\eta_i$
coefficients) and long distances effects, should be included.
Although strong corrections depend very weakly on $m_t$ in earlier
papers rather low values for $m_t$ were used~\cite{ajb3,gilman}. For
large top quark mass the $\eta_i$ coefficients were calculated in
Ref.~\cite{flynn}. For example, for $\Lambda_{QCD}=200$ MeV,
$m_c=1.7\,\mbox{GeV}/c^2$, $m_t=200\,\mbox{GeV}/c^2$ one has
$\eta_1=0.82$, $\eta_2=0.35$ and $\eta_3=0.62$ and we obtain $\Delta
m_K/m_K=6.40\times 10^{-15}$; for $\Lambda_{QCD}=250$ GeV,
$m_c=1.7\,\mbox{GeV}/c^2$, $m_t=200\,\mbox{GeV}/c^2$, $\eta_1=0.89$,
$\eta_2=0.62$ and $\eta_3=0.34$ we obtain $\Delta m_K/m_K=6.91\times
10^{-15}$. It means that in order to obtain a consistent value for
this mass difference $m_t>200,\mbox{ GeV}/c^2$.
This lower bound would be larger if we assumed values for
 $B_K$ smaller than $0.80$~\cite{ajb1}.

Our objective was just to put forward the narrow window for the
standard model parameters in processes like the $K_L$-$K_S$ mass
difference.
The main point, we would like to stress, in this
work is that the allowed values for $m_t$ from loop effects not
necessarily will coincide with the correct value of $m_t$ coming from
scattering processes and/or spectroscopy and in this case all
constraints in the physics beyond the standard model, coming from the
$\Delta m_K/m_K$, should be
revisited. That is, if the top quark mass is discovered with a mass
which do not fit the experimental value of $\Delta m_K/m_K$ and if
the improved knowledge of the mixing angles confirm the current
values then, it will imply that there is a new physics beyond the
standard model which contributes positively or negatively to the
$K_L$-$K_S$ mass difference. It means that this parameter instead of
constraints new physics may claim for it.

\acknowledgements

We would like to thank the
Con\-se\-lho Na\-cio\-nal de De\-sen\-vol\-vi\-men\-to Cien\-t\'\i
\-fi\-co e Tec\-no\-l\'o\-gi\-co (CNPq) for total (FP) and partial
(VP) financial support and C.O. Escobar and R. Zukanovich Funchal for
useful discussions.

\begin{figure}
{$\Delta m_K/m_K$ versus $m_t$ for the given values of the
phase $\delta$ and $m_c$. The dashed line
denotes the experimental value~\cite{pdg}.}
\label{}
\end{figure}

\begin{references}
\bibitem{fn} We use the notation of Ref.~\cite{pdg} but not for
the phase $\delta$.
\end{references}
\end{document}